# Differential Phase-Shift QKD in a 2:16-Split Lit PON with 19 Carrier-Grade Channels

N. Vokić, D. Milovančev, B. Schrenk, M. Hentschel, and H. Hübel

# Differential Phase-Shift QKD in a 2:16-Split Lit PON with 19 Carrier-Grade Channels

Nemanja Vokić, *Member, IEEE,* Dinka Milovančev, *Member, IEEE,*
Bernhard Schrenk, *Member, IEEE,* Michael Hentschel, and Hannes Hübel

*Abstract*— **We investigate the practical network integration of differential phase shift quantum key distribution following a cost-optimized deployment scheme where complexity is off-loaded to a centralized location. User terminal equipment for quantum state preparation at 1 GHz symbol rate is kept technologically lean through use of a directly-modulated laser as optical encoder. Integration in a passive optical network infrastructure is experimentally studied for legacy and modern optical access standards. We analyze the implications that result from Raman scattering arising from different spectral allocations of the classical channels in the O-, S-, C- and L-band, and prove that the quantum channel can co-exist with up to 19 classical channels of a fully-loaded modern access standard. Secure-key generation at a rate of 5.1×10⁻⁷ bits/pulse at a quantum bit error ratio of 3.28% is obtained over a 13.5 km reach, 2:16 split passive network configuration. The high power difference of 93.8 dB between launched classical and quantum signals in the lit access network leads to a low penalty of 0.52% in terms of error ratio.**

## I. Introduction

QUANTUM key distribution (QKD) provides the means for establishing a secret key between two communication parties by making use of fundamental laws of quantum mechanics. A wide range of QKD protocols have been proposed and experimentally implemented during the past years, of which some have made a significant leap towards commercialization. However, off-the-shelf QKD is mainly dedicated to secure point-to-point links against eavesdropping and is not optimized for cost-sensitive applications. On top of this, the practical network integration remains a challenge. While most network demonstrations have built on dark fiber and a combination of individual point-to-point links to form a meshed network [1-4], the co-existence of classical and

Manuscript received July 23, 2019; revised October 21, 2019 and November 6, 2019; accepted March 22, 2020. Date of publicationApril 1, 2020; date of current version April 21, 2020. This work was supported in part by funding from the European Union's Horizon 2020 research and innovation programme under grant agreement No 820474.

Nemanja Vokic, Dinka Milovancev, Bernhard Schrenk, Michael Hentschel, and Hannes Hübel are with the Austrian Institute of Technology, Center for Digital Safety&Security, Giefinggasse 4, 1210 Vienna, Austria (phone: +43 50550-4131; fax: -4150; e-mail: bernhard.schrenk@ait.ac.at).

Color versions of one or more of the figures in this article are available online at http://ieeexplore.ieee.org.

Digital Object Identifier 10.1109/JSTQE.2020.2983592

quantum channels in lit networks with carrier-grade power levels, zero-touch integration of the quantum channel such as required for brown-field networks, and operation over high optical network loss budgets has not been thoroughly investigated so far.

In this work we experimentally investigate a QKD integration scheme for passive optical networks (PON), which guarantees low capital and operational expenditures. By off-loading high-cost complex quantum-specific componentry to a centralized location such as a central office (CO), tail-end user QKD terminals can be kept relatively simple and cost-effective (Fig. 1). Differential phase shift keying (DPS) is chosen as QKD protocol which allows for an asymmetry in complexity between transmitter and receiver, thus supporting such an economic model. Moreover, optical expenditures are minimized through QKD operation in a lit (i.e., dark-fiber free) network with high robustness to several co-propagating classical optical signals. Raman scattering, a transmission impairment that is known for its wide spectral tails and its associated in-band crosstalk noise [5], is mitigated through a directional split between classical and quantum signals and narrowband optical filtering, similar as it is obtained through coherent optical reception [6, 7].

We will demonstrate secure-key generation at 5.1×10⁻⁷ bits/pulse and a quantum bit error ratio (QBER) of 3.28%, over a 13.5 km reach, 2:16 split PON by using a low-complexity, laser-based DPS transmitter at the optical network unit (ONU). Co-existence with GPON and NG-PON2 with up to 19 classical channels at a high power difference of up to 93.8 dB to the DPS-QKD channel is experimentally confirmed for a low QBER penalty of 0.52%.

The paper is organized as follows. Section II introduces the concepts for DPS-QKD and the network integration. Section III details the experimental setting for both application scenarios, including a characterization of signals and key elements. Section IV discusses the suppression of signal crosstalk and Raman noise, before Section V elaborates on the QKD performance with and without classical communication channels. Section VI finally concludes the work.

## II. DPS-QKD in Lit Passive Optical Networks

The practical implementation of quantum signals in lit fiber links requires a quantum channel that is robust to the loss conditions and the crosstalk noise that originates from the co-existence with classical channels. This challenge has been investigated in various ways for straight-line point-to-point links. Firstly by spacing the quantum channel spectrally far from the classical counterparts [8, 9], secondly by a reduced classical channel launch and allocation of the quantum channel in the Raman dip close to the optical carrier [10], and thirdly by exploiting space division multiplexing in specialized transmission fibers [11].

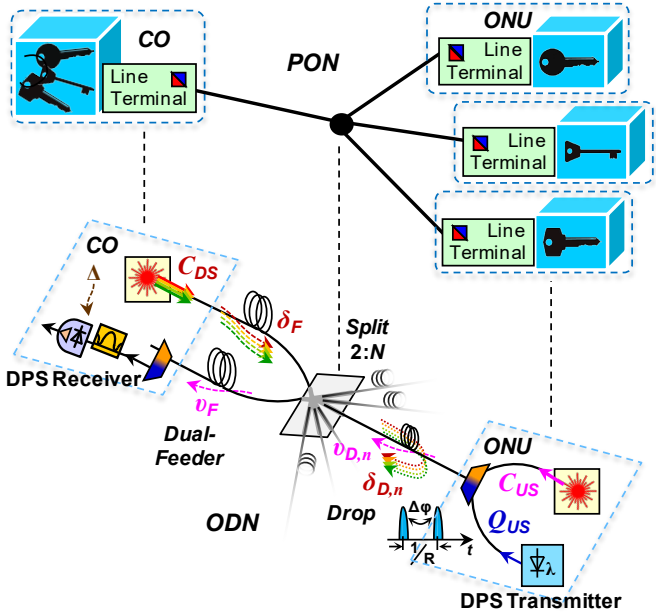

Fig. 1. Integration of DPS-QKD in a PON where high splitting loss is present and stimulated Raman scattering arises due to the classical channels.

For the particular case of PONs [12], the high optical loss budgets inherent to common tree architectures with filterless signal split and the strict compatibility with brown-field single-mode fiber deployments makes it impossible to bypass the splitting loss of an optical distribution network (ODN) through insertion of waveband filters. Moreover, the network is loaded with classical channels over a rather wide spectral range, which spans from the O- to the L-band around 1310 and 1590 nm, respectively.

Figure 1 illustrates such an application setting at which the quantum signal is integrated within an enhancement band towards wavelength-stacked classical signals. The directionless response of Raman scattering induced by the classical signals greatly determines the impact on the quantum channels and their integration approach in the bidirectional PON. Classical upstream signals $C_{US}$ originating at the ONUs will generally feature less lanes (i.e., 4 wavelengths for NG-PON2) than the downstream signals $C_{DS}$ from the CO. Additionally to wired communication, the advent of novel mobile network architectures and their integration with the wireline infrastructure leads to an increased number of downstream signals, as they for example apply for a wavelength division multiplexed (WDM) overlay of point-to-point links between the CO and remote antenna sites. In such a spectral setting, the quantum channel $Q_{US}$ is more likely to be implemented in the upstream direction.

In order to minimize the lightpath overlap with the downstream signals, a directional split can be implemented in the optical domain in virtue of the branching device at the PON tree: A dual-feeder ODN with 2:$N$ splitter avoids Raman noise $\delta_F$ at the dominant feeder length due to the high directivity of the tree splitter. The Raman noise contribution $\delta_D$, which arises at any of the $N$ drop fibers and which will eventually fall within the upstream direction, will have a comparably much lower magnitude due to the shorter drop length and the high splitting loss. To this end it shall also be recalled that although any downstream-induced crosstalk has to pass the splitter twice to reach the upstream receiver at the CO, the large number of $N$ drop fibers compensates for the double-pass for a single upstream splitting loss at all ports. Moreover, since classical upstream transmission is subject to time division multiple access (TDMA), the overall noise contribution over all $N$ drop fibers corresponds to a single continuous-mode signal, that nevertheless traverses the PON over its entire reach of drop ($v_D$) and feeder ($v_F$) fibers. Although the ODN loss attenuates Raman noise as it does for the quantum signals, the constant dark counts $\Delta$ at the single-photon avalanche photodetector (SPAD) of the CO introduce a limit for the compatible ODN loss budget.

The present investigation will target co-existence scenarios with different spectral configurations, corresponding to legacy and modern PON standards based on single and wavelength-stacked classical signal feed per transmission direction. The main aim is to demonstrate an economically viable path towards realizing QKD by successfully establishing key exchange at an arbitrary key rate, in contrary to high-performance QKD links that target a highest possible secure-key rate at arbitrary cost. For this purpose, DPS-QKD has been chosen. DPS-QKD encodes information in the optical phase difference $\Delta\varphi$ between two successive optical pulses [13]. The phase of an optical carrier is either modulated through the path choice in an unbalanced Mach-Zehnder interferometer or through an electro-optic phase modulator [14]. At the quantum receiver, the encoded optical phase is demodulated in a delay interferometer (DI) and measured by a single-photon detector. DPS features a simplistic reception scheme in only one measurement basis. It therefore utilizes the full rate of received photons and alleviates the QKD system from additional complexity due to measurement in two non-orthogonal settings and digital sifting, such as it is required for the BB84 protocol.

However, stable encoding and decoding in optical phase builds on interferometric elements at quantum transmitter and receiver [13], which need accurate and joint drift compensation. Alternative approaches to minimize the number of interferometric elements have been therefore proposed and include dual-laser schemes with optical injection [15] or direct electro-optic phase state preparation in an off-the-shelf distributed feedback (DFB) laser [16]. The latter exploits the chirp property of optical gain media [17] to enable direct phase modulation, previously reported for semiconductor optical amplifiers [18] and DFB lasers [19, 20]. Thermal tuning of the laser provides a degree of freedom to set different wavelength channels for DPS-QKD, which allows to share the centralized DI at the quantum receiver among multiple QKD transmitters. Bias current adjustment can be beneficially exploited for fine-tuning of the emission wavelength towards the spectral response of the shared DI. In this way an independent control is obtained for the spectral alignment of each QKD transmitter at the PON tree.

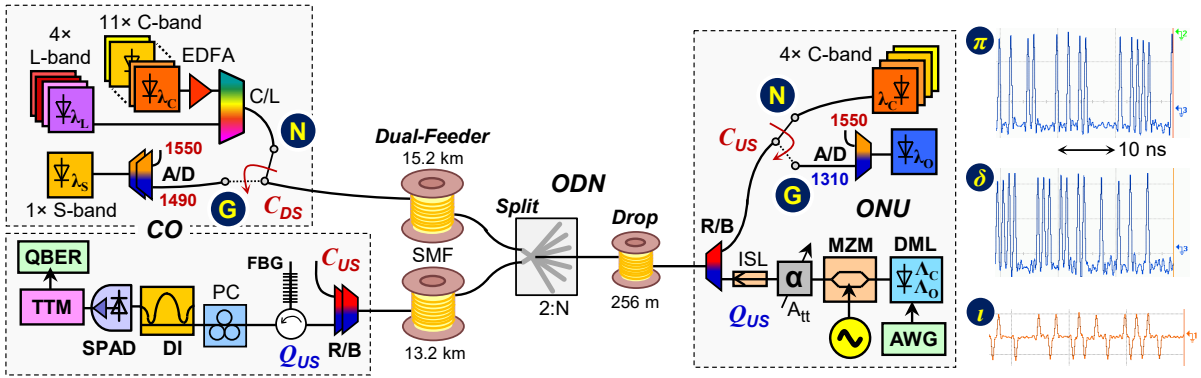

Fig. 2. Experimental setup for DPS-QKD transmission in a PON scenario. The insets show the DPS signal after reception, at high delivered optical power.

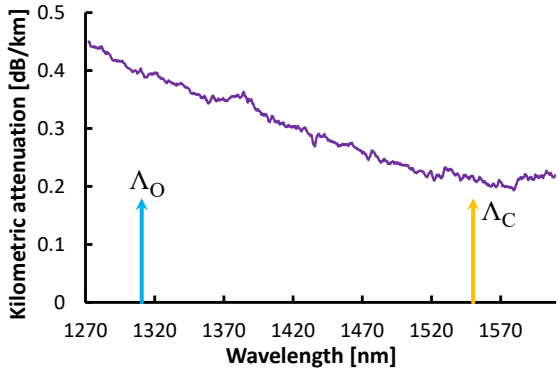

Fig. 3. Fiber attenuation at the two wavebands used for quantum signal transmission.

## III. EXPERIMENTAL SETUP AND SPECTRAL ALLOCATION

Figure 2 presents the experimental setup to investigate the co-existence of DPS-QKD with the classical access standards following the spectral setting of GPON [21] and NG-PON2 [22] in a simple power-splitting tree PON scenario.

### A. DPS transmitter

The DPS transmitter is located at the optical network unit (ONU) and launches the quantum signal in upstream direction. A butterfly-packaged directly modulated laser (DML) is employed to optically modulate the DPS signal at a symbol rate of R = 1Gbaud onto the quantum wavelength $\Lambda$. An arbitrary waveform generator (AWG) is applied for pseudo-random bit sequence generation at length $2^7$-1 and pulse shaping. The chirp property of the DML is exploited to obtain phase modulation. By tailoring the electrical drive signal according to the frequency modulation response of the laser, as shown for the electrical drive of the DML in the inset of Fig. 2 ($\iota$), direct phase modulation can be obtained. The use of chirp modulation in combination with a standard laser device alleviates the field-deployed transmitter from a bulky phase modulator, hence guaranteeing lowest cost when implementing the quantum transmitter at the user premises. Nevertheless, comparison is made to a DPS transmitter that builds on a LiNbO3 phase modulator. Moreover, pulse carving was applied to suppress the symbol edges, which would potentially allow a side channel attack. Since neither a 1310-nm externally modulated laser (EML) nor an O-band electro-absorption modulator were available for this purpose, a Mach-Zehnder modulator (MZM) was included at the DPS transmitter. However, an EML-based scheme for joint phase modulation and pulse carving has been recently demonstrated as a fully-integrated solution [16]. The DPS transmitter is completed by an optical attenuator that sets a mean photon number of $\mu = 0.1$ for the launched quantum signal, and an isolator that prevents external probing.

### B. DPS receiver

The corresponding DPS receiver is hosted at the CO, where it can be time-shared among multiple DPS transmitters in order to ensure lower capital expenditures of its more specific components by means of cost-sharing. It includes an optical phase demodulator comprised of a DI with an asymmetry that is adjusted to the symbol period of 1 ns in one of its arms. A manual polarization controller (PC) was used to optimize the response of the DI, which showed a polarization-dependent extinction. However, integrated DIs with polarization-independent characteristics have been demonstrated [23]. The demodulated optical DPS signal is shown as inset in Fig. 2 for a 1310-nm DPS transmitter that is based on either a LiNbO3 phase modulator ($\pi$) or the low-cost DML ($\delta$). A free-running InGaAs SPAD with a detection efficiency of 10% follows the optical demodulator. The events of the detector are registered by a time-tagging module (TTM) and are processed in real-time to estimate the raw key rate and the QBER.

### C. Optical distribution network

The transmission channel is given by the configuration of the ODN of the PON. This fiber plant between transmitter and receiver includes a dual-feeder fiber with 15.2 and 13.2 km in down- and upstream direction, respectively, a 2:N splitting stage that enables for a tree configuration, and a 256 m drop fiber span. Several fiber spans were based on ITU-T G.652.B compatible standard single-mode fiber (SMF). The characteristics for its kilometric loss are reported in Fig. 3. The average loss at 1550 and 1310 nm is 0.21 and 0.39 dB/km, respectively. There is no significant water peak. Although the higher losses at the O-band suggest to exclude this waveband for quantum signal transmission, the reduced Raman noise in combination with classical signals at the C-band renders it as superior.

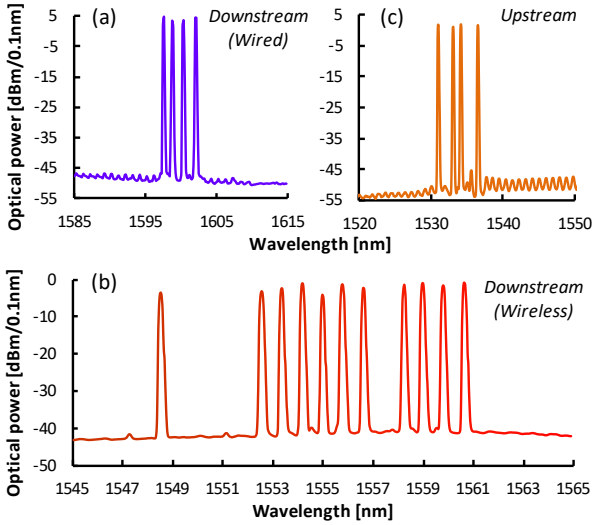

Fig. 4. Optical spectra for the classical channels for the NGPON-2 setting.

TABLE I
EXPERIMENTAL SETTING FOR CLASSICAL AND QUANTUM SIGNALS

| Signal | | Scenario / Characteristic | GPON | NG-PON2 | |
|---|---|---|---|---|---|
| **Downstream** | | Wavelength(s) [nm] | 1489 ($\lambda_S$) | 1597.62 1598.89 1600.17 1602.31 ($\lambda_L$) | 1548.51 to 1560.31 (11× $\lambda_C$) |
| | | Launch [dBm/λ] | 2.2 | 3.9 | -1.1 |
| **Upstream** | | Wavelength(s) [nm] | 1310 ($\lambda_O$) | 1531.12 1533.07 1534.25 1536.61 ($\lambda_C$) | |
| | | Launch [dBm/λ] | 0.3 | 1.6 | |
| **R/B** | | Filter passband [nm] | 1516 – 1650 | 1250 – 1410 | |
| | | Filter stopband [nm] | 1250 – 1516 | 1410 – 1650 | |
| **Quantum** | | Wavelength [nm] | 1550.12 ($\Lambda_C$) | 1310.55 ($\Lambda_O$) | |
| | | Launch [hν/symbol] | 0.1 | | |
| | | Narrowband optical filter bandwidth | DWDM A/D: 100 GHz | FBG: 14.6 GHz LAN-WDM: 800 GHz | |

### D. Classical channel load

The co-existence of the quantum channel ($Q_{US}$) with classical channels in a lit PON with shared fiber links has been evaluated under two prominent load conditions. Table I summarizes the chosen configurations used to launch classical signals in downstream ($C_{DS}$) and upstream ($C_{US}$) direction at the CO and ONU, which are based on the allocations found in GPON (G) and NG-PON2 (N) access networks. The first uses one S-band downstream signal at $\lambda_S$ = 1489 nm and one O-band upstream signal at $\lambda_O$ = 1310 nm. With this, a DPS configuration where the quantum signal is located at $\Lambda_C$ = 1550.12 nm is chosen. Optical add/drop (A/D) waveband filters after each of the classical transmitters are employed to clean optical out-of-band noise that would contaminate the quantum channel at the C-band. These additional filters can be seen as part of the mandatory WDM upgrade element that multiplexes the classical transmission and enhancement wavebands. The launched power for the single-wavelength classical channels were 2.2 and 0.3 dBm. Although these power levels are slightly below the maximum signal launch as foreseen in GPON, the mere fact that the typical loss budget of deployed PONs is in the order of 22 dB [24] therefore well below the GPON class B+ budget, makes this erosion of dynamic range due to a reduced launch permissible.

For the case of NG-PON2, wavelength-stacked signals in the C- and L-band apply for the classical down- and upstream, leaving the O-band as suitable enhancement waveband for quantum signal transmission at $\Lambda_O$ = 1310.55 nm. The optical spectra for the classical signals are presented in Fig. 4. At the

CO a set of four wavelengths from $\lambda_L$ = 1597.62 … 1602.31 nm (a) is utilized to emulate wired downstream transmission. A WDM overlay of 11 C-band wavelengths from 1548.51 … 1560.61 nm (b) is further loading the PON to account for wireless fronthauling, as it is considered for 5G scenarios. At the ONU the upstream is emulated by another set of four wavelengths from $\lambda_C$ = 1531.12 … 1536.61 nm (c). The average launched power for wired and wireless downstream channels were 3.9 and -1.1 dBm/λ, respectively, and the launch for the upstream was 1.6 dBm/λ. It shall be noted that continuous-mode classical upstream signals originating at a single ONU have been used rather than TDM signals of multiple ONUs distributed over the tree segment. However, given the summing of Raman contributions at the tree splitter, an equivalent – if not worse-case – emulation is provided with the actual experimental setup.

In order to protect the quantum channel from detrimental Raman noise, it is paramount to filter noise contributions in the optical excess bandwidth of the receiver. This task is accomplished by a cascade of two red/blue (R/B) waveband filters and a narrowband optical filter. For the NG-PON2 setting the latter is comprised of a fiber Bragg grating (FBG), which in case of O-band quantum signal transmission had a center wavelength at $\Lambda_O$ and a bandwidth of 14.6 GHz. Comparison will also be made to a commercial 800-GHz O-band filter following the ITU-T G.694.1 LAN-WDM standard. In the GPON scenario a 100-GHz DWDM A/D filter centered at $\Lambda_C$ is used as narrowband filter. Figure 5 shows the transmission functions of several optical narrowband filters relative to the wavelength of the quantum channel. The respective relative filter bandwidths, referenced to the center wavelength, reach from 6.4×10⁻⁵ for the FBG to 3.1×10⁻³ for the LAN-WDM filter.

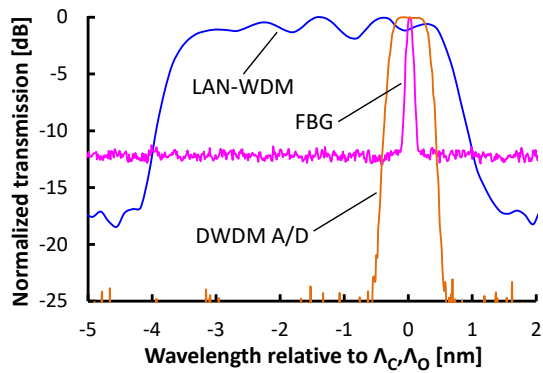

Fig. 5. Optical narrowband filters applied at the DPS receiver.

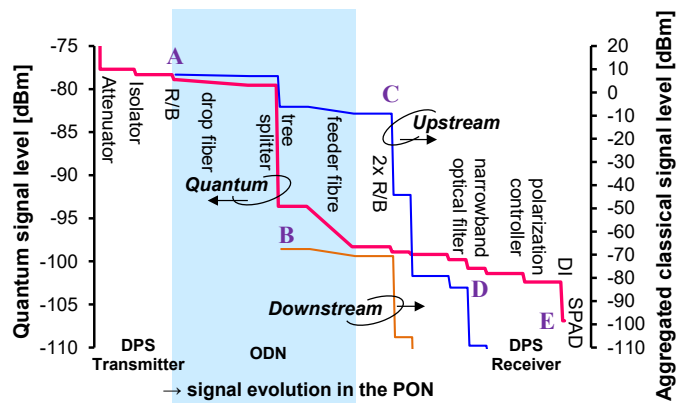

Fig. 6. Signal evolution for the NG-PON2 scenario with a 2:16 tree split.

## IV. CROSSTALK AND RAMAN NOISE

Communication systems that are based on direct photodetection require optical filters to select the target signal when more wavebands are jointly transmitted over the same channel. This ensures that out-of-band crosstalk is suppressed, which in case of QKD relates to the classical signals and their out-of-band Raman contribution. Moreover, narrow filtering can further minimize the effect of in-band Raman noise at the target waveband. In the following both spectral filtering mechanisms are being characterized.

### A. Signal crosstalk

Figure 6 shows the evolution of quantum and classical signal power level along the constituent elements of the lightpath. Data is provided for a PON with 2:16 split and QKD transmission at the O-band in co-existence with NG-PON2 down- and upstream. Several power values for the classical channels refer to the aggregated signal power rather than their Raman contribution, which is reported shortly in Fig. 7.

The co-propagating upstream is added to the channel with the R/B filter at the ONU (point A in Fig. 6) and fed together with the quantum signal over the entire ODN where it generates Raman noise. In contrary, the downstream does not contribute in the same manner since the dual-feeder architecture of the PON prevents a significant part of the downstream Raman noise. The latter is then limited to a much shorter drop fiber span located after passing through the high-loss passive branching device. However, the finite directivity of the power splitter leads to a feed of downstream signals in the upstream feeder direction (B). Even though this portion is rather small compared to the upstream, a close spectral spacing inherent to GPON, where the downstream is spaced by only 60 nm compared to the 240 nm displacement of the upstream, can lead to crosstalk for quantum signal reception. To avoid such crosstalk for any of the classical signals, coarse waveband filtering at the quantum receiver branch (C) ensures that out-of-band spectral content is sufficiently suppressed. The additional narrowband filter (D) contributes to this essential task, while also limiting the Raman contribution within the reception window to a minimum before single-photon detection (E).

### B. Raman noise

Since the in-band Raman contribution will be stronger than the Raman-inducing classical signals after filtering out-of-band crosstalk, an accurate characterization of these in-band components is required. Such a measurement was conducted using a single-photon optical spectrum analyzer (hν-OSA) consisting of a tunable grating filter and a SPAD with an efficiency of 10%. Figure 7 presents the Raman spectra for both, downstream and upstream direction. The spectra have been obtained at the drop fiber and upstream feeder fiber output, respectively. Measurement with the hν-OSA is performed after two R/B filters to reject the strong carrier components for down- and upstream.

Figure 7(a) shows the spectra for the GPON scenario with 2:16 split in the tree segment of the PON. The classical spectrum has been acquired in upstream direction, as it is intended for quantum signal transmission, and shows the strong upstream component at $\lambda_O$ with its material gain shoulder ($\alpha$). Although the downstream at $\lambda_S$ is counter-propagating, its carrier is visible due to an optical reflection at a flat connector of the upstream transmitter ($\rho$) and, in case that the upstream transmitter is disconnected, due to the finite directivity of the tree splitter ($\sigma$). The Raman scattering that originates from the optical carriers can only be seen through acquisition with the hν-OSA, for which results are reported at a resolution bandwidth of 1 nm up to the saturation point of its SPAD receiver. The Raman noise from the upstream (+) follows the spectral shape of the material gain shoulder ($\alpha$) and falls below the dark counts at a wavelength of 1430 nm or larger. The downstream, whose carrier is rather weak compared to the upstream, shows a marginal Raman contribution (×), which despite the close allocation towards the C-band vanishes at 1520 nm. The joint Raman noise for both, up- and downstream (◇), conveniently allows for a spectral allocation of the quantum channel in the C-band around 1550 nm.

For comparison, the Raman noise in downstream direction is also included in Fig. 7(a), as it would be perceived by a quantum receiver located at the ONU. The strong upstream signal introduces a Raman contribution at the drop fiber (●) without attenuation of either carrier or Raman noise by any splitting loss. This contribution is orders of magnitudes higher than in upstream direction and extends towards the L-band, however, the upstream signal was launched in continuous rather than in burst mode. Nevertheless, this represents a worst-case scenario with a low user subscription rate at the PON tree. Moreover, the contribution of the much closer downstream signal is now strongly visible in the C-band. The downstream-induced Raman noise (▲) entirely saturates the hv-OSA receiver and shows wide spectral wings reaching from 1330 nm to beyond the L-band. Needless to say, the joint contribution of down- and upstream (■) renders any quantum transmission as impossible.

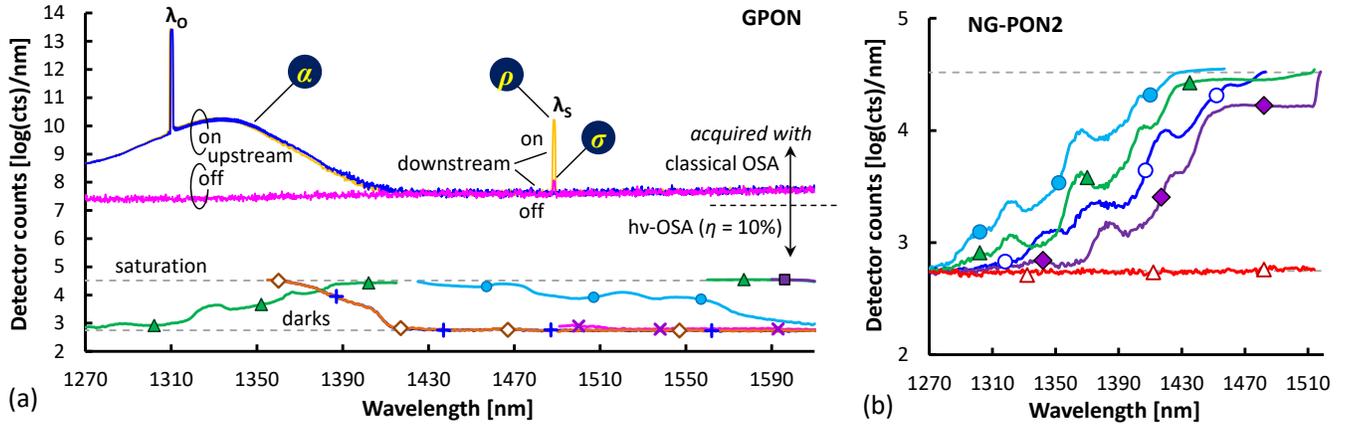

Fig. 7.  Signal and Raman spectra for 2:16 split PON, induced by (a) GPON and (b) NG-PON2 transmission.

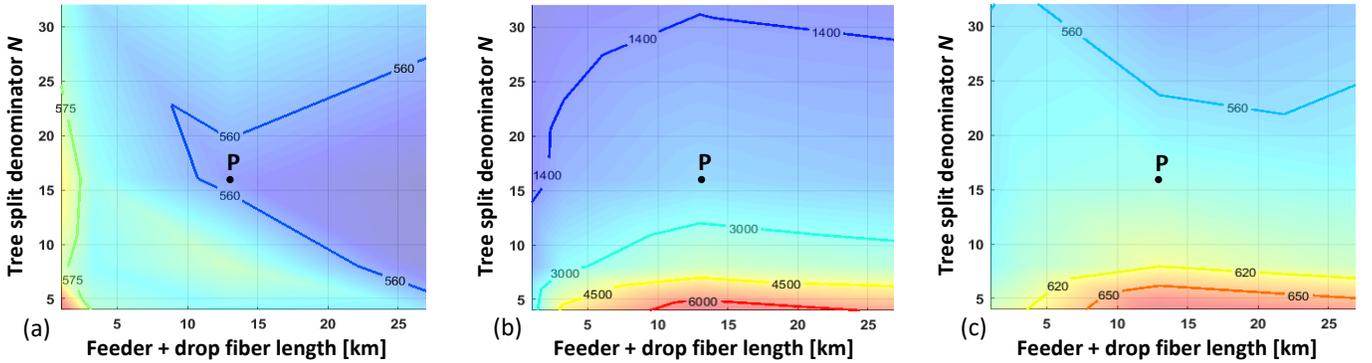

Fig. 8.  Detector counts registered by the quantum receiver for (a) downstream-induced Raman noise with LAN-WDM filter, and upstream-induced noise with (b) LAN-WDM and (c) FBG narrowband optical filter.

In case of NG-PON2 several classical signals are allocated to the C- and L-band. The corresponding Raman spectra that are induced by various combinations of classical signals are presented in Fig. 7(b) for the O-band region at which quantum signal transmission is considered. In downstream direction, the Raman contributions from all (●), wired upstream, wired downstream and WDM overlay due to wireless fronthauling, lead to a significant noise contribution at the desired quantum channel wavelength $\Lambda_O$. In case that the slightly closer allocated upstream in the 1530 to 1540 nm region is switched off (○) the noise is just marginally above the dark count rate of the hv-OSA receiver. For the given launch power of the WDM overlay, its Raman contribution is the least critical (◆). The acquired Raman noise in upstream direction confirms that this direction is the preferred option for the practical implementation of a quantum channel: With all Raman-inducing sources being powered on (▲), the overall contribution is weaker than in downstream direction. It can also be noticed that all downstream signals together (△) have no impact on the quantum signal reception, as it is similar to the GPON scenario.

The wide spectral wings of the Raman contributions underpin the necessity of the narrowband optical filter, which reduces the effective reception bandwidth to a fraction of the quantum waveband. For example, the narrow FBG bandwidth improves the rejection of Raman noise by ~23 dB with respect to CWDM filtering. The Raman counts that are aggregated over the reception bandwidth and received by the actual quantum receiver at the CO, including optical filters, DI and SPAD, are presented in Fig. 8 for the NG-PON2 scenario. For this measurement the tree split denominator $N$ and the feeder fiber length at fixed drop length have been varied in order to investigate the impact of ODN parameters on the Raman limits that are imposed for the reception of the quantum signal. Reported values include ~520 dark counts/s of the SPAD receiver.

Figure 8(a) shows the downstream-induced Raman counts with LAN-WDM filter at the quantum receiver. There is just a marginal contribution above the dark-count level, which rises slightly for short feeder length and low split ratio since for these conditions the downstream injection for the drop span, which generates the downstream Raman contribution, is maximized.

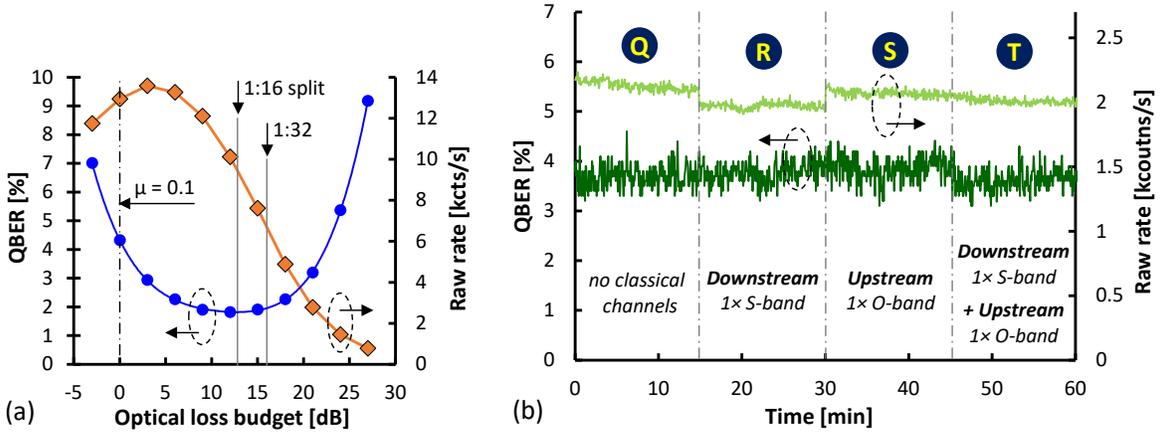

Fig. 9. (a) Back-to-back QKD performance in dark PON in NG-PON2 scenario with DML-based QKD transmitter. (b) QKD performance in lit GPON setting.

The upstream-induced Raman noise is reported in Fig. 8(b) and 8(c) for a narrowband filter based on LAN-WDM and FBG, respectively. The concentrated splitting loss determines the generated Raman noise since the classical pump for the feeder section and the Raman contribution of the drop fiber are both attenuated. Raman noise is then pronounced for longer (>5 km) fiber spans. A saturation effect can be noticed at ~15 km, which is explained by the loss that is introduced for longer fiber spans. For the broader LAN-WDM filter, a peak Raman contribution of ~1730 c/s can be anticipated for a 2:16 split after subtracting the constant dark count rate. This high value can be significantly reduced to ~60 c/s through filtering with the spectrally narrow FBG passband.

## V. QKD PERFORMANCE

### A. Dark PON

The QKD performance was first evaluated for a dark-channel scenario for which the fiber span has been further emulated by equivalent attenuation. In order to align with the measurements for which the classical signals are present, temporal slicing of the received DPS signal at a width of 30% of the symbol period has been applied, as will be explained shortly. The results for this back-to-back evaluation, which serves as a reference, are presented in Fig. 9(a) as raw key rate (◆) and QBER (●) over the optical loss budget between ONU and CO. Measurements are shown for the DML-based DPS transmitter in the NG-PON2 application setting, for which a raw key rate of 3.4 kb/s and a QBER of 2.9% can be obtained for a loss budget of 20 dB. Even for a high budget of 26 dB, a rate of 1 kb/s can be obtained. The lowest QBER found is 1.82% at a budget of 12 dB, where the constant dark counts $\Delta$ of the SPAD are standing in a smaller ratio to the received photon counts. This QBER is clearly below the 5% threshold for which a secure key can be generated [25], with an estimated fraction of 46% or ~4.6 kb/s of secure key to remain after error correction and privacy amplification at this loss budget. This corresponds to a secure-key generation rate of $3.6 \times 10^{-7}$ bits/pulse. A 1% penalty in QBER is experienced in the loss budget range from 3.5 to 19.8 dB, meaning a large dynamic range of 16.3 dB. Saturation effects at a lower optical budget lead to SPAD saturation and increase the QBER due to after-pulsing. Compared to the DPS transmitter based on phase modulator and to the GPON application setting, minor differences apply for these back-to-back results.

### B. Lit PON

Finally, the DPS QKD performance has been evaluated in terms of raw key rate and QBER for transmission over a 2:16 split, 13.5 km reach PON (point P in Fig. 8) in presence of classical channels. Temporal filtering within 30% of the symbol period is applied to suppress Raman noise between the carved DPS pulses. Figure 9(b) reports the performance for the GPON scenario. Reference is made to 1550-nm DPS transmission without classical channels (Q), for which a raw key rate of 2.12 kb/s and a QBER of 3.69% ($3\sigma = 0.15\%$) is obtained for the given loss PON budget. The QBER is still below the key generation threshold and a secure key of ~360 bits/s is estimated to remain, corresponding to a rate of $3.6 \times 10^{-7}$ secure bits/pulse. There is no degradation in QBER when switching on either the downstream S-band channel (R), the upstream O-band channel (S), or both classical channels (T). This proves the robustness of the quantum channel in co-existence with classical PON signals.

The QKD performance for the NG-PON2 scenario with wavelength-stacked classical signals is presented in Fig. 10. Measurements in Fig. 10(a) build on a DSP transmitter with LiNbO$_3$ based phase modulator and a quantum receiver with LAN-WDM filter for in-band Raman noise suppression. Without classical channels (A), a raw key rate of 3.25 kb/s and a QBER of 3.16% is obtained. In presence of the 15 downstream channels (B) the QBER raises just marginally by 0.26%. However, the addition of the four upstream channels (C) in the more critical co-propagation direction results in a QBER excursion to 7.55%. The strong Raman noise can also be noticed in the registered counts. This underpins the importance of a narrow optical reception filter. Figure 10(b) shows the same measurements using the FBG as Raman suppression filter. The raw key rate is slightly lower due to higher insertion loss for the FBG-based filter that involves a double pass through a circulator. Co-existence with the four wired (D) and the 11 WDM overlay channels (E) in downstream direction shows a QBER degradation less than 0.1%. More importantly, the impact of the four upstream channels (F) reduces to a small QBER excursion of 0.38%. With all co-existing classical channels in both, down- and upstream direction (G), the QBER worsens by 0.52%. This proves that QKD can be successfully integrated in a lit PON infrastructure with multi-channel classical transmission.

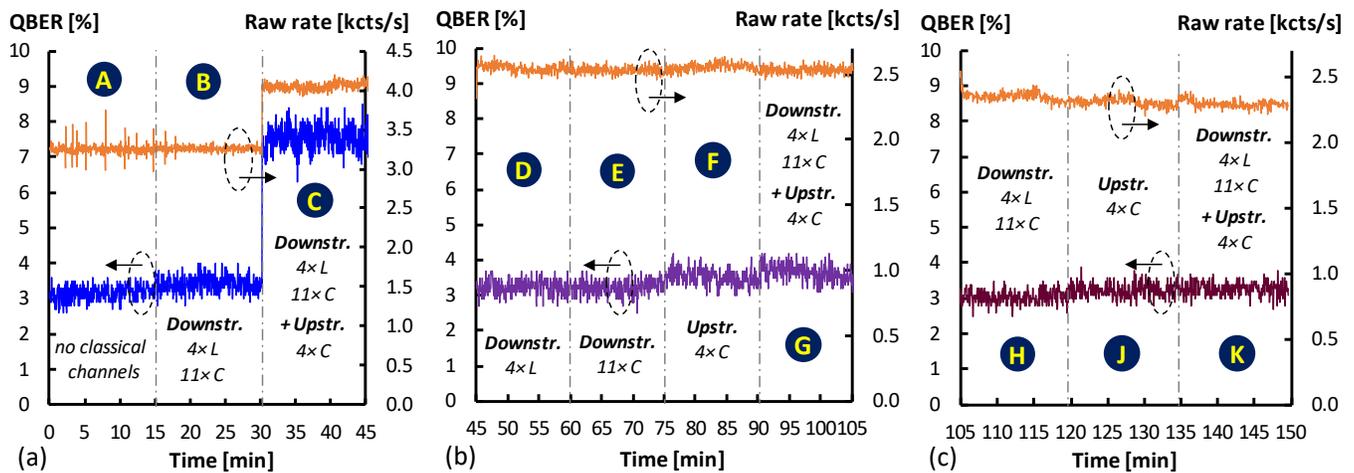

Fig. 10. QKD performance in lit NG-PON2 setting for (a) DPS transmitter with external phase modulator and LAN-WDM filter at the receiver, (b) after migration to the FBG filter at the receiver, and (c) after switching to the DML-based DPS transmitter.

To also prove QKD integration with DML-based DPS transmitter, Fig. 10(c) reports the performance under co-existence with all downstream (H), all upstream (J) and all classical channels together (K). The raw key rate and QBER for the last case are 2.29 kb/s and 3.28%, respectively, with a minor degradation due to present classical channels and similar to the results obtained for the external phase modulator. As such it proves the PON integration of QKD with cost-optimized user terminal equipment.

## VI. CONCLUSION

The practical integration of QKD in a PON infrastructure has been experimentally studied for GPON and NG-PON2 centric wavelength allocations. The noise that results from Raman scattering of the classical access signals suggests to implement the QKD link in upstream direction. DPS-QKD has been chosen as protocol due to its potential to realize transmitting and receiving subsystems with high asymmetry in terms of complexity. To keep the user terminal technologically lean and thus cost-effective, the more complex quantum receiver is deployed as shared detector at the CO, while a rather simplistic quantum encoder based on a DML resides at the ONUs. For such a DPS-QKD system, a raw key rate of 10.1 kb/s and a QBER of 1.82% is obtained at an optical loss budget of 12 dB. Network operation of the DPS-QKD in a 13.5 km reach, 2:16 split PON yields a secure-key rate of 0.51 kb/s at a QBER of 3.28%. Robustness to Raman noise is gained through narrowband optical filtering. In case of NG-PON2 with up to 19 classical channels in the C- and L-band, the QBER penalty for O-band QKD remains as low as 0.52%. No penalty with respect to external phase modulation has been observed due to the low-cost, DML-based QKD transmitter.

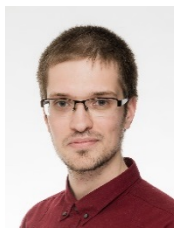

**Nemanja Vokić** (M'19) was born in Serbia in 1990. He received the M.Sc. degree in electrical and computer engineering from the University of Novi Sad, Serbia in 2014. He obtained his Ph.D. degree from Vienna University of Technology in 2018. For his Ph.D. work, he designed electronic integrated circuits for transmitters and receivers used for optical communications. He was then a post-doc at Vienna University of Technology, designing mm-wave ASICs for 5G remote-radio-heads. In 2019, he joined AIT Austrian Institute of Technology, Vienna. His research interests include design of analog and RF integrated circuits, opto-electronic circuit integration, coherent optical communication for telecom, datacom and quantum communications.

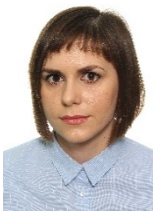

**Dinka Milovančev** (M'19) was born in Serbia in 1990. She finished her M.Sc. studies in 2014 in Microelectronics from the University of Novi Sad, Serbia. She did her Ph.D. research at Vienna University of Technology in Austria at the Institute of Electrodynamics, Microwave and Circuit Engineering from 2014 up to 2019. Her Ph.D. research interests include circuit design of integrated avalanche photodiode receivers and their applications for optical wireless and fiber communication. Since 2019 she is employed at AIT Austrian Institute of technology, Vienna, where she works in the field of coherent optical fiber communication in classical as well as quantum optical domain.

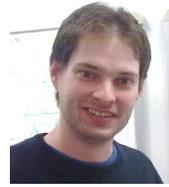

**Bernhard Schrenk** (S'10-M'11) was born 1982 in Austria and received the M.Sc. ('07) degree in microelectronics from the Technical University of Vienna. He was at the Institute of Experimental Physics of Prof. A. Zeilinger, where he was involved in the realization of a first commercial prototype for a quantum cryptography system, within the European SECOQC project. From 2007 to early 2011 he obtained his Ph.D degree at UPC BarcelonaTech, Spain. His Ph.D thesis on multi-functional optical network units for next-generation Fiber-to-the-Home access networks was carried out within the FP7 SARDANA and EURO-FOS projects. In 2011 he joined the Photonic Communications Research Laboratory at NTUA, Athens, as post-doctoral researcher and established his research activities on coherent FTTH under the umbrella of the FP7 GALACTICO project. In 2013 he established his own research force on photonic communications at AIT Austrian Institute of Technology, Vienna, where he is working towards next-generation metro-access-5G networks, photonics integration technologies and quantum optics.

Dr. Schrenk has authored and co-authored ~140 publications in top-of-the-line (IEEE, OSA) journals and presentations in the most prestigious and highly competitive optical fiber technology conferences. He was further awarded with the Photonics21 Student Innovation Award and the Euro-Fos Student Research Award for his PhD thesis, honoring not only his R&D work but also its relevance for the photonics industry. He was elected as Board-of-Stakeholder member of the Photonics21 European Technology Platform in 2017. During his extensive research activities he was and is still engaged in several European projects such as SARDANA, BONE, BOOM, APACHE, GALACTICO, EURO-FOS and the Quantum Flagship project UNIQORN. In 2013 he received the European Marie-Curie Integration Grant WARP-5. In 2018 he was awarded by the European Research Council with the ERC Starting Grant COYOTE, which envisions coherent optics everywhere.

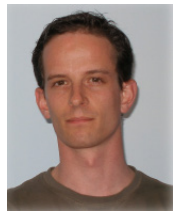

**Michael Hentschel** was born 1975. He received his master's degree in electrical engineering in 1998 from the Vienna University of Technology and obtained his PhD in the field of laser technology in 2001. To this end he worked on the optimization of femtosecond laser amplifiers and the generation of attosecond X-ray pulses. After 2 years of further research at the Photonics Institute at the Vienna University of Technology, he spent 3 years in an Austrian laser company with R & D tasks in the field of ultrashort pulse lasers. In 2006 he began his work on Quantum Information and Quantum Cryptography at the University of Vienna and the Austrian Academy of Sciences with Prof. Anton Zeilinger. He has specialized in the development of sources of entangled photons and their use in quantum cryptography systems. Since the summer of 2009 he is an employee of the AIT Austrian Institute of Technology and has been working on various national and European projects involving quantum cryptography.

Dr. Hentschel has authored and co-authored ~70 publications in peer reviewed journals and presentations in international conferences.

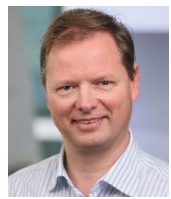

**Hannes Hübel** obtained his PhD in 2004 from Queen Mary, University of London, UK. In the same year he joined the quantum information group at the University of Vienna, headed by Anton Zeilinger, to work on experimental quantum communication, in particular Quantum Key Distribution (QKD). In 2008, he presented the first realization of a fully automated QKD system based on entanglement, within the European SECOQC project. He then worked as a post-doctoral researcher at the University of Waterloo, Canada, focusing on experimental demonstrations of multipartite entanglement. In 2010, he became assistant professor at the University of Stockholm, Sweden. Since 2015, he leads the experimental QKD development at the AIT Austrian Institute of Technology in Vienna, Austria.

Dr. Hübel has authored and co-authored ~50 publications in top-of-the-line journals and more than 30 invited and contributed presentations at international conferences. He builds on practical experience from the coordination of national (TransQ, QKD Telco, KVQ) and contracted industrial projects, as well as on experience in steering and participation in large European projects including QT-Flagship projects (QAP, SECOQC, UNIQORN, CIVIQ).